\documentclass[epj]{svjour}
\usepackage{graphics}
\begin{document}
\title{Study of strongly nonlinear oscillators using the Aboodh transform and the homotopy perturbation method}
\author{K. Manimegalai\inst{1} \and Sagar Zephania C F\inst{1} \and P. K. Bera\inst{2} \and P. Bera\inst{3} \and S. K. Das\inst{4} \and Tapas Sil\inst{1}  
\thanks{\emph{Present address:} Department of Physics, Indian Institute of Information Technology Design and Manufacturing Kancheepuram, Chennai-600127, Tamil Nadu, India} %
}                     
\offprints{\email{tapassil@iiitdm.ac.in}}          
\institute{Department of Physics, Indian Institute of Information Technology Design and Manufacturing Kancheepuram, Chennai-600127, Tamil Nadu, India  \and Department of Physics, Dumkal College, Basantapur, Dumkal, Murshidabad-742303,West Bengal,  India \and School of Electronics Engineering, VIT University, Vellore-623014, Tamil Nadu, India \and Department of Mechanical Engineering,IIT Ropar, Rupnagar-140001, Punjab, India 
}
\date{Received: date / Revised version: date}
%
\abstract{
	 A generalized equation is constructed for a class of classical oscillators with strong anharmonicity  which are not exactly solvable.  Aboodh transform based homotopy perturbation method (ATHPM) is applied to get the approximate analytical solution for the generalized equation and hence some physically relevant anharmonic oscillators are studied as the special cases of this solution.
ATHPM is very simple and hence provides the approximate analytical solution of the generalized equation without any mathematical rigor. The solution from this simple method not only shows excellent agreement with the exact numerical results but also found to be better accuracy in comparison to  the solutions obtained from other established approximation methods whenever compared for physically relevant special cases.
%
\PACS{31.15.xp \and 43.40.Ga}
} 
\maketitle
\section{Introduction}
\label{intro}
Most of the physical systems are nonlinear in nature and hence they are mostly not exactly solvable
\cite{bonham1966use,bender1968analytic,nayfehperturbation}. Although, getting numerical solution for the differential equations representing systems involving nonlinearity are sometimes easy, one desires to get the analytic solution of such problems as they carry more information and hence give a better insight into the system. Perturbation method 
is a widely used method for finding an approximate solution to complex nonlinear systems, especially with the nonlinear term appears as an additional term of small order to an exactly solvable problem.
As the  equations for many nonlinear systems do not have small parameter, application of perturbation technique is highly restricted. 
There are many techniques for solving nonlinear oscillator problems analytically such as 
the harmonic balance method \cite{nayfehnonlinear}, the Krylov-Bogolyubov-Mitropolsky method \cite{bogoli͡ubov1961asymptotic}, weighted linearization method \cite{agrwal1985weighted}, perturbation procedure for limit cycle analysis \cite{chen1991perturbation}, modified Lindstedt-Poincare method \cite{cheung1991modified}, Adomain decomposition method \cite{adomian1988review}, artificial parameter method \cite{liu1997new}, Homotopy Analysis method (HAM) \cite{liao1992proposed,liu2013symbolic} and so on. Most of these methods are not only involved the calculational rigor but also failed to handle problems with strong nonlinearity properly. 
Energy balance method (EBM) proposed by J. H.  He \cite{he2002preliminary} based on the variational principle,  is one of the commonly used non-perturbative techniques. This heuristic approach is found to be working well for several strongly nonlinear systems \cite{ganji2009periodic,he2003determination,mehdipour2010application}. 
There exists  another non-perturbative analytic  method due to He \cite{he2008improved} 
known as frequency-amplitude-formulation (FAF) which finds a lot of successful applications \cite{ganji2009periodic,langari2011,elnaggar2012applications}. 
FAF does not require a small parameter and a linear term in the differential equation. 
Recently, Nofal {\it et. al}, \cite{nofal2013analytical} employed FAF followed by EBM, to study some physically relevant anharmonic oscillators with strong anharmonicities and concluded that this FAF-EBM method has given much better accuracy in comparison to that obtained by using EBM alone.  

J. H. He developed the homotopy perturbation method (HPM) for solving linear, nonlinear, initial and boundary value problems \cite{he1999homotopy,he2000coupling}. In this method, the solution is given in an infinite series usually converging to an accurate solution \cite{yildirim2009homotopy,biazar2015}. HPM is found to be very efficient in solving problems with strong nonlinearity in classical \cite{he1999homotopy,he2000coupling,biazar2011new} as well as quantum mechanical domain \cite{bera2012homotopy}. 

Aboodh introduced a transform \cite{aboodh2013new} derived from the classical Fourier integral for solving ordinary and partial differential equations easily in the time $(t)$ domain. Aboodh transform (AT) has been applied for different types of problems and is found to be a very simple technique 
to solve differential  equations.\\
We construct a generalized nonlinear differential equation which, under certain approximation,
reduces to different physically relevant problems, such as, vibration of tapered beam,
motion of a particle in arranged parabola, Mathews-Lakshmanan oscillator, etc.  
Aboodh transform based homotopy perturbation method (ATHPM) is applied to find out a generalized solution to these problems and hence  to get the displacement $(x)$ and the frequency  oscillation $(\omega)$ for the special cases. We compare ATHPM results to those obtained from FAF-EBM and exact numerical calculations (RK4) to check its accuracy.

This paper is organized as follows. In section \ref{sec:2}, we demonstrate briefly the formulation of ATHPM. Applications of ATHPM to study some physically relevant anharmonic oscillators have been shown in section  \ref{sec:3}. Finally, in section  \ref{sec:4} we provide a brief discussion and our conclusions.
\section{Formalism}
\label{sec:2}
If $x(t)$ is the piecewise continuous function  of $t$, the corresponding Aboodh transform  is defined as, \cite {aboodh2013new} 
\begin{equation}\label{aboodh2}
A[x(t)]=x(\nu)=\frac{1}{\nu}\int_{0}^{\infty}x(t)e^{-\nu {t}} dt,\nu \exists(k_1,k_2),
\end{equation}
where, $k_1,k_2>0$ and may be finite or infinite.
Some properties of Aboodh transform necessary for our calculation are as follows,
\begin{eqnarray}\label{aboodh3}
A[x{''}(t)]&=&\nu^2x(\nu)-\frac{x{'}(0)}{\nu}-x(0)\nonumber\\
A[\cos{at}]&=&\frac{1}{{\nu^2}+{a^2}},\nonumber\\
A[t\sin{at}]&=&\frac{2a}{(\nu^2+a^2)^2},\\
A[t^n]&=&\frac{n!}{\nu^{n+2}}\nonumber
\end{eqnarray}
Let us consider a nonlinear inhomogeneous differential equation as,
\begin{equation}\label{nleq1}
Lx(t)+\omega^2x(t)+Rx(t)+Nx(t)=g(t),
\end{equation}
with the initial conditions at $t=0, x(0)=a$ and $x{'}(0)=0$
Here, $L$ is the second order linear differential operator ($L\equiv\frac{d^2}{dt^2}$), $R$ is the linear operator having an order less than $L$,   $N$  is the nonlinear operator, $g(t)$ is the inhomogeneous term and $\omega^2$ is a parameter. 
Now, taking the Aboodh transform on both sides of eq.(\ref{nleq1}), we get,
\begin{equation}\label{anleq1}
A[Lx(t)]+\omega^2A[x(t)]+A[Rx(t)]+A[Nx(t)]=A[g(t)].
\end{equation}
Using the differential properties of the Aboodh transform (AT) as mentioned above and the initial conditions, eq.(\ref{anleq1}) can be written as,
\begin{eqnarray}\label{a2nleq1}
x(\nu)&=&\left(\frac{1}{\nu^2+\omega^2}\right)x(0)+\frac{x{'}(0)}{\nu(\nu^2+\omega^2)}-\left({\frac{1}{\nu^2+\omega^2}}\right)A[Rx(t)]\nonumber\\
&&\quad-\left(\frac{1}{\nu^2+\omega^2}\right)A[Nx(t)]-\left(\frac{1}{\nu^2+\omega^2}\right)A[g(t)].
\end{eqnarray}
Taking the inverse Aboodh transform on both sides of eq.(\ref{a2nleq1}), we get,
\begin{eqnarray}\label{ainleq1}
x(t)&=&X_0(t)-A^{-1}\left[\left(\frac{1}{\nu^2+\omega^2}\right)A[Rx(t)]\right]\nonumber \\
&&-A^{-1}\left[\left(\frac{1}{\nu^2+\omega^2}\right)A[Nx(t)]\right]-A^{-1}\left[\left(\frac{1}{\nu^2+\omega^2}\right)A[g(t)]\right]
\end{eqnarray}
where,
\begin{equation}\label{upnleq1}
X_0(t)=\left(\frac{1}{\nu^2+\omega^2}\right)x(0)+\frac{x{'}(0)}{\nu(\nu^2+\omega^2)}
\end{equation}
According to the homotopy perturbation method \cite{he1999homotopy,he2008improved,bera2012homotopy}, we may expand $x(t)$ in power of an embedded parameter $p$  $(0 \leq p \leq 1)$ as, $x(t)=\sum_{n=0}^{\infty}p^nx_n(t)$ and nonlinear term $Nx(t)=\sum_{n=0}^{\infty}p^nH_n(x)$, where He's polynomial $H_n(x)$ can be written as,
\begin{equation}\label{hepoly}
H_n(x)=\frac{1}{n!}\frac{d^n}{dp^n}\left[N\sum_{n=0}^{\infty}p^nx_n(t)\right],\hspace{0.15in} n=0,1,2,3,...
\end{equation}
By the construction of homotopy, here we get an exactly solvable problem for $p=0$ whereas $p=1$ corresponds to the nonliear problem 
for which we are trying to find the solution.
Applying HPM and substituting the value of $x(t)$ and $Nx(t)$ in eq.(\ref{ainleq1}) in terms of the power series of $p$ and $H_n(x)$, we get,

\begin{eqnarray}\label{hpmnleq1}
\sum_{n=0}^{\infty}p^nx_n(t)&=&X_0(t)-p\left(A^{-1}\left[\left(\frac{1}{\nu^2+\omega^2}\right)A\left[R\sum_{n=0}^{\infty}p^nx_n(t)\right]\right]\right. \nonumber \\
&&+A^{-1}\left[\left(\frac{1}{\nu^2+\omega^2}\right)A\left[\sum_{n=0}^{\infty}p^nH_n(t)\right]\right]  \nonumber \\
&&\left.+A^{-1}\left[\left(\frac{1}{\nu^2+\omega^2}\right)A[g(t)]\right]\right).
\end{eqnarray}
Comparing the coefficient of like power of $p$ on both sides, we get  the following relations from eq.(\ref{hpmnleq1}), 
\begin{eqnarray}\label{c1nleq1}
p^0:x_0(t)&=&X_0(t),\\
p^1:x_1(t)&=&-A^{-1}\left[\left(\frac{1}{\nu^2+\omega^2}\right)A\left[Rx_0(t)\right]\right]-A^{-1}\left[\left(\frac{1}{\nu^2+\omega^2}\right)A[H_0(x_0(t))]\right]\nonumber \\
&&-A^{-1}\left[\left(\frac{1}{\nu^2+\omega^2}\right)A[g(t)]\right].
\end{eqnarray}
\begin{eqnarray}\label{c3nleq1}
p^2:x_2(t)&=&-A^{-1}\left[\left(\frac{1}{\nu^2+\omega^2}\right)A[Rx_1(t)]\right]-A^{-1}\left[\left(\frac{1}{\nu^2+\omega^2}\right)A[H_1(x_1(t))]\right]\nonumber \\
&&-A^{-1}\left[\left(\frac{1}{\nu^2+\omega^2}\right)A[g(t)]\right].
\end{eqnarray}
The approximate solution, as $p\rightarrow 1$, is,
\begin{equation}\label{apxsolnleq1}
x(t)=\lim\limits_{p\to1}\sum_{n=0}^{\infty}p^nx_n(t)=x_0(t)+x_1(t)+x_2(t)+x_3(t)+.....
\end{equation}
Here, $x_0(t)$ is the zeroth order term which corresponds to the solution $p=0$ homotopy ie, the exactly solvable part of the equation. The first order correction is represented by $x_1$ and $x_2$ is the second order term and so on. It is to be noted that the approximate solution of $x(t)$ in eq.\ref{apxsolnleq1} is independent of the expansion parameter $p$ or any other perturbative parameter. The HPM solution not only converges very fast but also gives the exact solution with the certain assumption \cite{biazar2015,ghorbani2009beyond}.
\section{Applications}
\label{sec:3}
We construct the following differential equation representing the general form of a group of nonlinear oscillators which are profusely used for describing physical systems \cite{mehdipour2010application,nofal2013analytical,he2008comment,davodi2009application,zhang2009periodic} encountered in science and engineering as,
\begin{equation}\label{geq}
\frac{d^2x}{dt^2}+\frac{\lambda{x}+a_1x(\frac{dx}{dt})^2+a_2x^3\left(\frac{dx}{dt}\right)^2+a_3x^3+a_4x^5}{1+b_1x^2+b_2x^4}=0,
\end{equation}
where, $\lambda,a_1,a_2,a_3,a_4,b_1$ and $b_2$ are arbitrary parameters. Let us rewrite eq.(\ref{geq}) as,
\begin{eqnarray}\label{geq1}
\frac{d^2x}{dt^2}+\omega^2x&=&(\omega^2-\lambda)x-b_1x^2\frac{d^2x}{dt^2}-b_2x^4\frac{d^2x}{dt^2}-a_1x\left(\frac{dx}{dt}\right)^2\nonumber \\
&&-a_2x^3\left(\frac{dx}{dt}\right)^2-a_3x^3-a_4x^5.
\end{eqnarray}
We apply AT on both sides of eq.(\ref{geq1}) to get,
\begin{eqnarray}\label{ageq1}
x(\nu)&=&\left(\frac{1}{\nu^2+\omega^2}\right)x(0)+\frac{x{'}(0)}{\nu(\nu^2+\omega^2)}+(\omega^2-\lambda)\left(\frac{1}{\nu^2+\omega^2}\right)A[x]\nonumber \\
&&-b_1\left(\frac{1}{\nu^2+\omega^2}\right)A\left[x^2\frac{d^2x}{dt^2}\right]-b_2\left(\frac{1}{\nu^2+\omega^2}\right)A\left[x^4\frac{d^2x}{dt^2}\right]\nonumber \\
&&-a_1\left(\frac{1}{\nu^2+\omega^2}\right)A\left[x\left(\frac{dx}{dt}\right)^2\right]-a_2\left(\frac{1}{\nu^2+\omega^2}\right)A\left[x^3\left(\frac{dx}{dt}\right)^2\right]\nonumber \\
&&-a_3\left(\frac{1}{\nu^2+\omega^2}\right)A[x^3]-a_4\left(\frac{1}{\nu^2+\omega^2}\right)A[x^5].
\end{eqnarray}
Taking inverse AT on both sides of eq.(\ref{ageq1}) and applying the initial conditions, at $t=0$,  $x(0)=a$ and $x{'}(0)=0$, we obtain,

\begin{eqnarray}\label{aigeq1}
x(t)&=&a\cos\omega{t}+(\omega^2-\lambda)A^{-1}\left[\left(\frac{1}{\nu^2+\omega^2}\right)A[x]\right]\nonumber\\
&&-b_1A^{-1}\left[\left(\frac{1}{\nu^2+\omega^2}\right)A\left[x^2\frac{d^2x}{dt^2}\right]\right]-b_2A^{-1}\left[\left(\frac{1}{\nu^2+\omega^2}\right)A\left[x^4\frac{d^2x}{dt^2}\right]\right]\nonumber\\
&&-a_1A^{-1}\left[\left(\frac{1}{\nu^2+\omega^2}\right)A\left[x\left(\frac{dx}{dt}\right)^2\right]\right]\nonumber \\ &&-a_2A^{-1}\left[\left(\frac{1}{\nu^2+\omega^2}\right)A\left[x^3\left(\frac{dx}{dt}\right)^2\right]\right]\nonumber\\
&&-a_3A^{-1}\left[\left(\frac{1}{\nu^2+\omega^2}\right)A[x^3]\right]-a_4A^{-1}\left[\left(\frac{1}{\nu^2+\omega^2}\right)A[x^5]\right].
\end{eqnarray}
With the help of the properties of AT and inverse AT, we obtain the coefficients of $p^0$ and $p^1$  from eq.(\ref{c1nleq1}) as follows,
\begin{eqnarray}\label{c4geq1}
p^0:x_0(t)&=&a\cos(\omega {t}),\nonumber\\
p^1:x_1(t)&=&\frac{1}{2\omega}\left[a(\omega^2-\lambda)+\frac{3}{4}b_1 a^3 \omega^2+\frac{5}{8}b_2a^5\omega^2-\frac{1}{4}a_1a^3\omega^2-\frac{1}{8}a_2a^5\omega^2\right.\nonumber\\
&&\left. -\frac{3}{4}a_3 a^3-\frac{5}{8}a_4 a^5\right] t sin{\omega}t+\frac{1}{8\omega^2}\left[\frac{1}{4}\left(a_1+b_1\right)a^3\omega^2\right.\nonumber\\
&&\left. +\frac{1}{16}\left(a_2+5b_2\right)a^5\omega^2-\frac{1}{4}a_3 a^3-\frac{5}{16}a_4 a^5\right] (cos{\omega}t-cos3{\omega}t)\nonumber\\
&&+\frac{1}{24\omega^2}\left[\frac{1}{16}\left(a_2+b_2\right)a^5\omega^2
-\frac{1}{16}a_4 a^5\right] (cos{\omega}t-cos5{\omega}t).
\end{eqnarray}
To avoid the secular term, we put the coefficient of $t sin\omega t$ equal to zero, i.e.,
\begin{equation}\label{secular1}
a(\omega^2-\lambda)+\frac{3}{4}b_1 a^3 \omega^2+\frac{5}{8}b_2a^5\omega^2-\frac{1}{4}a_1a^3\omega^2-\frac{1}{8}a_2a^5\omega^2-\frac{3}{4}a_3 a^3-\frac{5}{8}a_4 a^5=0,
\end{equation}
which gives the angular frequency of nonlinear oscillation as, 
\begin{equation}\label{frege1}
\omega=\sqrt{\frac{8\lambda+6a_3a^2+5a_4a^4}{8+2a^2(3b_1-a_1)+(5b_2-a_2)a^4}}.
\end{equation}
Using eq.(\ref{secular1}) in eq.(\ref{c4geq1}), we get the analytic solution from eq.(\ref{aigeq1}) to the generalized equation eq.(\ref{geq}), considering the first order approximation as,
\begin{eqnarray}\label{apxsolgeq1} 
x_{ATHPM}(t)&=& a\cos\omega {t} \nonumber \\
&&+\frac{1}{8\omega^2}\left[\frac{1}{4}\left(a_1+b_1\right)a^3\omega^2+\frac{1}{16}\left(a_2+5b_2\right)a^5\omega^2-\frac{1}{4}a_3 a^3\right.\left. -\frac{5}{16}a_4 a^5\right]\nonumber \\ 
&& \times (cos{\omega}t-cos3{\omega}t)\nonumber \\
&&+\frac{1}{24\omega^2}\left[\frac{1}{16}\left(a_2+b_2\right)a^5\omega^2\right.\left. -\frac{1}{16}a_4 a^5\right] (cos{\omega}t-cos5{\omega}t).
\end{eqnarray}

We shall study different physically relevant cases considering different sets of force parameters in eq.(\ref{geq}). 
\subsection{Case 1: Motion of a particle on a rotating parabola}
We consider the equation of motion of a particle sliding down freely on a parabola which is rotating about its axis \cite{nayfehnonlinear,marinca2006application}, 
\begin{equation}\label{peq}
\frac{d^2x}{dt^2}+\frac{\omega_0^2{x}+4q^2x(\frac{dx}{dt})^2}{1+4q^2 x^2}=0,
\end{equation}
which also represents the movement of the double-slider mechanism \cite{ganji2012energy}. This may be obtained from the generalized equation eq.(\ref{geq}) by choosing, $\lambda=\omega_0^2,a_1=4q^2,a_2=a_3=a_4=0,b_1=4q^2$ and $b_2=0$. 
The frequency of this nonlinear oscillator by employing ATHPM may be obtained  from eq.(\ref{frege1}) substituting the parameters as mentioned above and can be written as,
\begin{equation}\label{pfeq}
\omega=\frac{\omega_0}{\sqrt{1+2q^2 a^2}},
\end{equation}
which is the same as given by Davodi et al \cite{davodi2009application} obtained using the amplitude frequency formulation and also the same given by Nofal et. al \cite{nofal2013analytical} employing frequency amplitude formulation  based energy balance method (FAF-EBM). The approximate solution of eq.(\ref{peq}) is obtained by ATHPM from eq.(\ref{apxsolgeq1}) as,
\begin{equation}\label{pxh}
x_{ATHPM}(t)=a\cos{\omega t}+\frac{1}{4}q^2a^3\left(cos{\omega t}-cos3{\omega t}\right).
\end{equation}
Also, the approximate result by the FAF-EBM is,
\begin{equation}\label{pxf}
x_{FAF-EBM}(t)=a\cos\left(\frac{\omega_0}{\sqrt{1+2q^2 a^2}}t\right).
\end{equation}
\begin{figure}
	\centering
	\resizebox{0.9\textwidth}{!}{\includegraphics{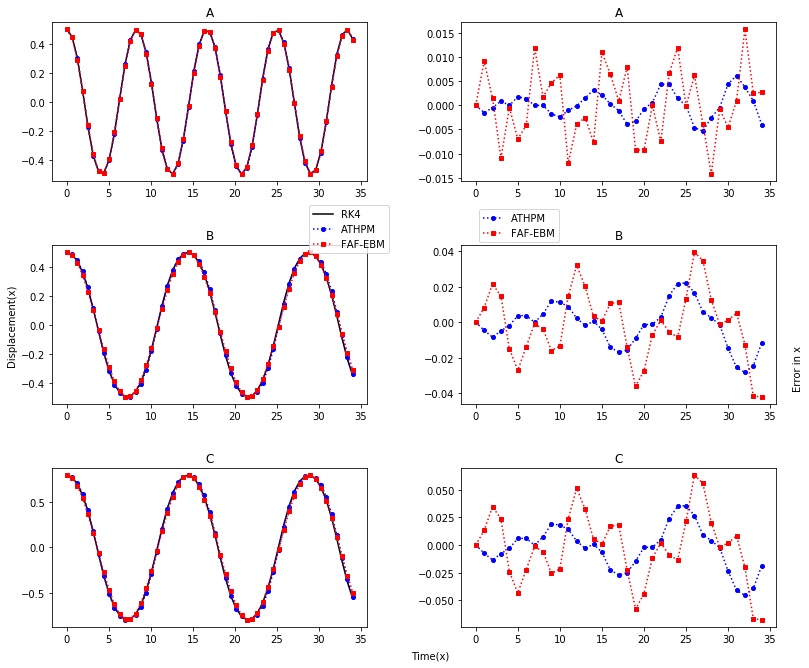}}
	\caption{Plot of variation of displacements $x_{RK4}(t)$ (black solid line), $x_{ATHPM}(t)$  (blue circles), and $x_{FAF-EBM}(t)$ (red squares) with time ($t$) are shown in the three panels of the left column for three parameter sets A, B, C in the top panel, middle panel, and bottom panel respectively. Errors in approximate calculations with respect to RK4, $\epsilon_{xA}$ and $\epsilon_{xF}$ for the same parameter sets are displayed in the right column.}
	\label{figpara1}       
\end{figure}
We plot the displacement obtained from ATHPM $x_{ATHPM}(t)$ (blue circles) eq.(\ref{pxh}) with  increasing time $t$ for three sets of values of parameters ($a,\omega_0 ,q$) in the left column of Figure \ref{figpara1} and compared with the same given by FAF-EBM method $x_{FAF-EBM}(t)$ (red squares) eq.(\ref{pxf}) and also that obtained by numerical solution of the  eq.(\ref{peq}) employing forth order Runge-Kutta (RK4) method  $x_{RK4}(t)$ (black solid line). It is seen that for all the parameter sets 
[A:(0.5,0.8,0.5) for top panel, B:(0.5,0.5,0.8) middle panel and C:(0.8,0.5,0.5) in the bottom panel], approximate displacements match extremely well with the $x_{RK4}$. 


The error in approximate solutions of  displacement  with respect to its values calculated using RK4, $\epsilon_{xA}(=x_{RK4}-x_{ATHPM},$ blue circles) and $\epsilon_{xF}(=x_{RK4}-x_{FAF-EBM},$ red squares) are displayed in the right column for the same parameter sets. Errors involved in both of the approximate solutions of $x(t)$ are small (maximum value  $\epsilon_{xA}~0.045$ and  $\epsilon_{xF}~0.074$) within the ranges of the time $t$ and for the  parameters considered. 
All three panels in the right column show that accuracy of $x_{ATHPM}(t)$ is much improved in comparison to $x_{FAF-EBM}(t)$.


\begin{figure}
	\centering
	\resizebox{0.9\textwidth}{!}{\includegraphics{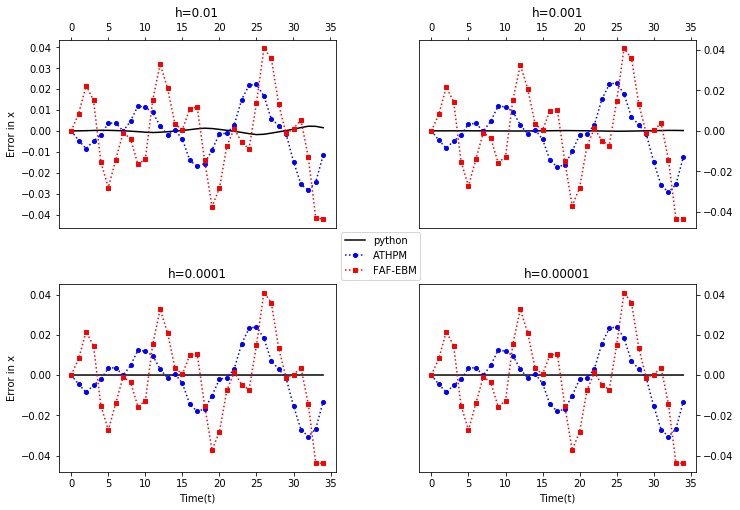}}
	\caption{Plot of  error in  displacement  obtained from Python function odeint $\epsilon_{xP}$ (black solid line), eq.(\ref{pxh}) $\epsilon_{xA}$ (blue circles) and eq.(\ref{pxf}) $\epsilon_{xF}$ (red squares)  with respect to those calculated from RK4 versus time for different values of the mesh size  ($h$) of time $t$.}
	\label{mathematic2}   
\end{figure}
In  Figure \ref{figpara1}, a comparison of the results obtained from ATHPM and FAF-EBM is done considering those  from RK4 method as the reference. In order to check the reliability of RK4 results, we compute the displacement  $(x_{PY})$ by solving eq.\ref{peq} using Python (function `odeint') for the parameter set $B$. We plot in Figure \ref{mathematic2}, the error $\epsilon_{xP}(=x_{RK4}-x_{PY}$, along with  $\epsilon_{xA}$ and  $\epsilon_{xF}$ as a function of time $t$ for different values of mesh size ($h ~ 0.01, 0.001, 0.0001$ and $0.00001$) of time $t$.  It is observed that $\epsilon_{xP}$, (black solid line) remains  very close to zero, through out the span of time considered here, whereas $\epsilon_{xA}$, (blue circles) and  $\epsilon_{xF}$, (red squares) having maximum error $0.028$ and $0.046$ respectively. The errors   $\epsilon_{xA}$ and  $\epsilon_{xF}$ remains the same for all values of $h$. This gives the confidence about the accuracy of the numerical solutions using RK4 which is taken as the reference when we compare solutions from two approximation methods such as ATHPM and FAF-EBM.      

\begin{table}\label{tabErrPar}
	\begin{center}
		\caption{Comparison of maximum error in calculation of displacement ($x(t)$) in ATHPM ($\epsilon^{max}_{xA}$) and FAF-EBM ($\epsilon^{max}_{xF}$) methods for different values of the force parameters. The locations of occurrence of maximum errors for both models are displayed in last four columns.}
		\begin{tabular}{|c|c|c|c|c|c|c|}
			\hline
			parameters & $\epsilon^{max}_{xA}$ & $\epsilon^{max}_{xF}$ & 
			$t^{max}_A$ & $x^{max}_A$ & $t^{max}_F$ & 
			$x^{max}_F $\\
			\hline
			$a=0.5,\omega_0 =0.5$ & & & & & &\\
			\hline
			$q=0.2$ &	0.0011 &   0.0003 & 20.0 & -0.4451  & 19.6 & -0.4803\\
			\hline
			0.5 &  0.0025 & 0.0025  &     9.4  & -0.0011    & 9.4 & -0.0012\\
			\hline
			0.8 &	-0.0033 & -0.0296 & 20.0 &-0.4038  & 20.0 & -0.4038\\
			\hline
			1.0 &	-0.0259 & -0.0603 & 20.0  & -0.2114 & 20.0 & -0.2114\\
			\hline
			1.5	& 0.08643 & 0.1031 & 12.0 & -0.2980 & 14.7 & 0.2747\\  
			\hline
			\hline
			$a=0.5,q=0.5$ & & & & & &\\
			$\omega_0 =0.2$ &	0.0026 & -0.0061 &  20.0 & -0.4108 & 20.0 & -0.4107\\
			\hline
			0.5 &  0.0025 &  0.0025  &     9.4 &  -0.0012   & 9.4 & -0.0012\\
			\hline
			0.8 &	-0.0039 & -0.0127 & 20.0 & -0.4166  & 20.0 & -0.4166\\
			\hline
			1.0 & 0.0119 & 0.0207 & 18.2 & -0.0313  & 18.8 & 0.2575\\
			\hline
			1.5 &	0.0002 & 0.0002 & 20.0 & -0.4997 & 20.0 & 0.4997\\
			\hline
			\hline
			$\omega_0=0.5, q=0.5$ & & & & & &\\
			$a=0.2$ &	0.0004 & 0.0001 & 20.0 &-0.1780 &  19.6 & -0.1921\\
			\hline
			0.5 &	0.0025 &  0.0025  &     9.4 &  -0.0012    & 9.4 &  -0.0012\\
			\hline
			0.8 &  -0.0052 & -0.0473 & 20.0 & -0.6461 &  20.0 & -0.6461\\
			\hline
			1.0 &  -0.0518 & -0.1205 & 20.0 & -0.4228 &  20.0 & -0.4228\\
			\hline
			1.5 &  0.2593 & 0.3095 & 12.0 & -0.8942  & 14.7 &  0.8243\\
			\hline
		\end{tabular}
		
	\end{center}
\end{table}
In Table~\ref{tabErrPar}, we display, the maximum error $\epsilon^{max}_{xA}$ in displacement obtained from ATHPM in the second column, the occurrence of  maximum error at the time    ($t^{max}_A$) in the fourth column and the values of the displacement $x^{max}_A$ at $t^{max}_A$ in the fifth column for a range of values of the force parameters.  We have also displayed the values of the same
quantities obtained from FAF-EBM  $\epsilon^{max}_{xF}$, $t^{max}_F$, and  $x^{max}_F$  in the third, sixth and seventh columns respectively to compare the corresponding ATHPM results. We see, the order of magnitude of the error are same in the displacement obtained from both the approximate methods although the ATHPM results are  found to give  better numerical accuracy than FAF-EBM specially at  larger values of the force parameters.

\subsection{Case 2: Tapered Beam}
\begin{figure}\label{fig3}
	\centering
	\resizebox{0.75\textwidth}{!}{\includegraphics{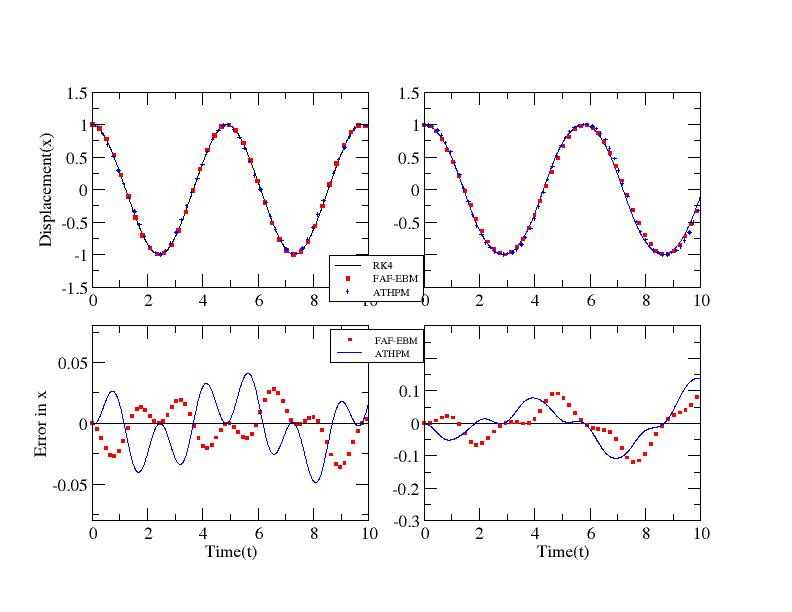}}
	\caption{ Plot of variation of displacements $x_{RK4}(t)$ (solid line), $x_{ATHPM}(t)$ (circles), and $x_{FAF-EBM}(t)$ (squares) with time $t$ are shown in the top row for two parameter sets $U$, and $V$ in the left panel, and the right panel respectively. Errors in approximate calculations $\epsilon_{xA}$ and $\epsilon_{xF}$ for the same parameter sets are displayed in the bottom panels.}
	\label{fig:2taper1}       
\end{figure}

Tapered members are increasingly used in the construction industry because of their unique ability to combine efficiency, economy and aesthetics -- the three corner stones of structural art \cite{billinton1985}. Tapered beam is an important model for engineering structures having variable stiffness along the length such as tree-branches, turbine blades, bridges etc. Fundamental vibration mode of a tapered beam can be expressed as the following nonlinear differential equation \cite{akbarzade2012dynamic,hoseini2009large,gorman1975free} 
\begin{equation}\label{tbeq}
\frac{d^2x}{dt^2}+\frac{x+\varepsilon x{\left(\frac{dx}{dt}\right)}^2+\beta x^3}{1+\varepsilon x^2}=0.
\end{equation}
The same equation may be obtained by choosing the arbitrary parameters in eq.(\ref{geq}) as $\lambda=1, a_1=\varepsilon,a_2=0,a_3=\beta,a_4=0,b_1=\varepsilon,b_2=0$. The ATHPM frequency can be obtained from eq.\ref{frege1} as given below
\begin{equation}\label{fretb}
\omega=\sqrt{\frac{4+3\beta a^2}{4+2\varepsilon a^2}}, 
\end{equation}
which is the same as given by FAF-EBM method \cite{nofal2013analytical}. The approximate solution by ATHPM $(x_{ATHPM})$ to eq.(\ref{tbeq}) obtained from eq.(\ref{apxsolgeq1}) is as follows,
\begin{equation}
x_{ATHPM}(t)=a\cos\omega t+\frac{1}{8\omega^2}\left[\frac{1}{2}\varepsilon a^2\omega^2-\frac{1}{4}\beta a^3\right](\cos\omega t-\cos3\omega t, 
\end{equation}
where the same given by FAF-EBM \cite{nofal2013analytical} is,
\begin{equation}
x_{FAF-EBM}(t)=a\cos\left(\sqrt{\frac{4+3\beta a^2}{4+2\varepsilon a^2}t}.\right).
\end{equation}
In Figure ~\ref{fig:2taper1}, we plot the exact displacement from numerical solution  
$x_{RK4}$ (solid line), $x_{FAF-EBM}$ (squares) and $x_{ATHPM}$ (circles)
for two different parameter sets $U$($a=1, \varepsilon=0.1, \beta=1$, in the upper left panel) and $V$($a=1, \varepsilon=1, \beta=1$, in the upper right panel). We have also compared the variation of errors $\epsilon_{xA}$ and $\epsilon_{xF}$ with time for the aforementioned parameter set in corresponding bottom panels. 

It is found from the top panels of the Figure \ref{fig:2taper1}, that the approximate solutions for the displacement of tapered beam mimic with those obtained from the RK4 very well for the range of time and parameter sets considered for this study. In this case, the accuracies of the solutions obtained by ATHPM and FAF-EBM are similar. A close look at the error-graphs $\epsilon_{xA}$($=x_{RK4}-x_{ATHPM}$, solid line) and $\epsilon_{xF}$($=x_{RK4}-x_{FAF-EBM}$, squares) verses time $t$, displayed in the bottom panels of the figure corroborates the conclusion made from the plots presented in the top panels.

\subsection{Case 3: Autonomous Conservative Oscillator}
Let us consider the force parameters, $a_1=\varepsilon, a_2=2\alpha, a_3=\beta, a_4=\gamma, b_1=\varepsilon$ and $b_2=\alpha$.  From eq.(\ref{geq1}) we obtained the equation of motion as,
\begin{equation}\label{acoeq}
\frac{d^2x}{dt^2}+\frac{\lambda x+\varepsilon x{\left(\frac{dx}{dt}\right)}^2+2\alpha x^3{\left(\frac{dx}{dt}\right)}^2+\beta x^3+\gamma x^5}{1+\varepsilon x^2+\alpha x^4}=0,
\end{equation}
which represents the free vibrations of an autonomous conservative oscillator with  fifth order nonlinearities \cite{mehdipour2010application,hamdan1997large,chen2009application}. Here motion is assumed to start from the position of maximum displacement with zero initial velocity. The parameter $\lambda$ is an integer which may take values from $-1,0,1$ and $\varepsilon, \alpha, \beta, \gamma$ are positive parameters. The solution to the above equation may be readily obtained from the generalized solutions eq.(\ref{frege1}), and eq.(\ref{apxsolgeq1}) with the help of ATHPM. The approximate frequency as a function of amplitude is obtained as, 
\begin{equation}\label{acoFreh}
\omega=\sqrt{\frac{8\lambda+6\beta a^2+5\gamma a^4}{8+4\varepsilon a^2+3\alpha a^4}},
\end{equation} 
which is the same as given by the FAF-EBM  method \cite{nofal2013analytical} but differs from the  expression of frequency reported by Mehdipour et.al \cite{mehdipour2010application} using EBM,

\begin{equation}\label{acoFebm}
\omega_{EBM}=\frac{1}{\sqrt{3}}\sqrt{\frac{12\lambda+9\beta a^2+7\gamma a^4}{8+4\varepsilon a^2+\alpha a^4}}.
\end{equation}
The ATHPM solution of eq.(\ref{acoeq}) is, 

\begin{eqnarray}\label{acoxh}
x_{ATHPM}(t)&=&a\cos\omega t+\frac{1}{8\omega^2}\left[\frac{1}{2}\varepsilon a^3\omega^2+\frac{7}{16}\alpha a^5\omega^2-\frac{1}{4}\beta a^3-\frac{5}{16}\gamma a^5\right] \nonumber \\ &&\times (\cos\omega t-\cos3\omega t)\nonumber\\
&&+\frac{1}{24\omega^2}\left[\frac{3}{16}\alpha a^5\omega^2-\frac{1}{16}\gamma a^5\right](\cos\omega t-\cos5\omega t),
\end{eqnarray}
where, the same given by FAF-EBM as,
\begin{equation}\label{acoxf}
x_{FAF-EBM}(t)=a\cos \omega t.
\end{equation}
It is noted that unlike $x_{FAF-EBM}$, $x_{ATHPM}$ contains terms from higher harmonics ($3\omega$ and $5\omega$).\\
\begin{figure}\label{fig4}
	\centering
	\resizebox{0.75\textwidth}{!}{\includegraphics{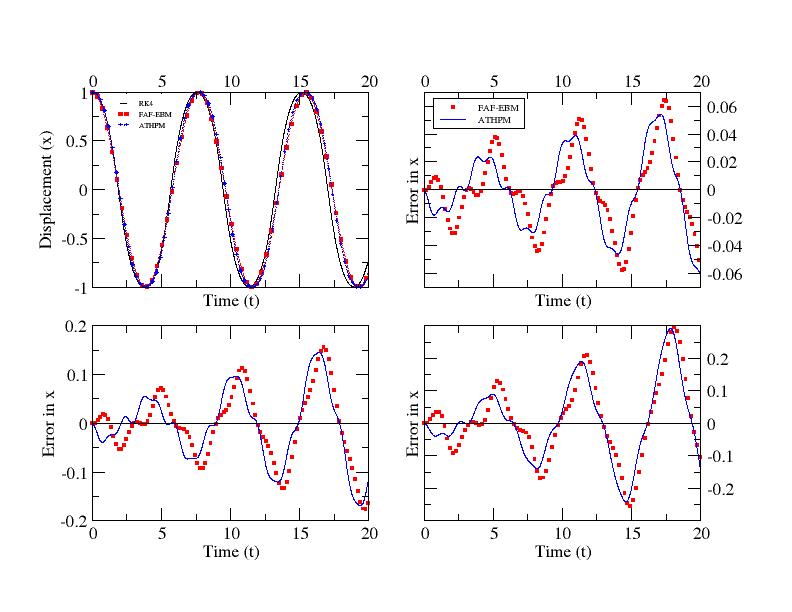}}
	\caption{The plots of displacements $x_{ATHPM}(t)$ (circles), $x_{FAF-EBM}(t)$ (squares), and $x_{RK4}(t)$ (solid line) as a function of time $t$ for  parameter set $P$ is displayed in the left top panel. Comparison of errors    $\epsilon_{xA}$ (solid line) and $\epsilon_{xF}$  (squares) are given for parameter sets $Q$ (right top panel), $R$ (left bottom panel) and $S$ (right bottom panel).}
	\label{fig:3aco1}       
\end{figure}
The variation of displacement obtained from ATHPM, $x_{ATHPM}(t)$ (circles) with time $t$ for of values of parameter set $P$ ($a=1,\epsilon=0.2, \alpha=0.2, \beta=0.1, \gamma=0.1$) in the left top panel of Figure \ref{fig:3aco1} and compared with the same given by FAF-EBM method $x_{FAF-EBM}(t)$ (squares) eq.(\ref{pxf}) and also that obtained by numerical solution of the  eq.(\ref{peq}) employing fourth order Runge-Kutta (RK4) method $x_{RK4}(t)$ (solid line). We have also compared the variation of errors $\epsilon_{xA}$ (solid line) and $\epsilon_{xF}$  (squares) within the time range 0 to 20, for the parameter sets $Q$ (1.0,0.5,0.5,0.3,0.2) right top panel, $R$ (1.0,1.0,0.5,0.3,0.2) left bottom panel and $S$ (1.0,1.5,0.2,0.2,0.1) right bottom panel.

It is seen from the left top panel that all three curves match very well at initial stage the approximate solutions start deviating slowly from the corresponding exact values $x_{RK4}(t)$ as the time increases. The same is observed in error graphs shown in the other three panels. 

\subsection{Case 4: Mathews and Lakshmanan Oscillator}
Mathews and Lakshmanan \cite{mathews1974unique} presented a nonlinear system which obeys equation of motion as follows, 
\begin{equation}\label{MLeq}
\frac{d^2x}{dt^2}+\frac{\alpha^2x\mp kx{\left(\frac{dx}{dt}\right)}^2}{1+kx^2}=0
\end{equation}
This equation of motion was obtained from the Lagrangian density 
for a relativistic scalar field which arises in the context of the theory of  elementary particle. 
Eq.(\ref{MLeq}) is a simpler form of the general equation in eq.(\ref{geq}).
Considering the arbitrary parameters as  $\lambda=\alpha^2, a_1=\pm k, a_2=a_3=a_4=0, b_1=\mp k$ and $b_2=0$ one can arrive at eq.(\ref{MLeq}).  
The frequency of the nonlinear oscillator which is obtained by ATHPM from eq.(\ref{frege1}) as,
\begin{equation}\label{moF}
\omega_{ATHPM}=\frac{\alpha}{\sqrt{1\pm ka^2}},
\end{equation}
which is the same as the exact frequency \cite{mathews1974unique}. We obtain first order correction term as, $x_1=0$.  Thus, the displacement in ATHPM is, 
\begin{equation}\label{mox}
x_{ATHPM}=acos \left[ \frac{\alpha}{\sqrt{1\pm ka^2}} t \right].
\end{equation}
Therefore, we get the exact solution by ATHPM of the Mathews and Lakshmanan nonlinear oscillator.
\section{Conclusion}
\label{sec:4}
A generalized equation is constructed which reduces to   
strongly nonlinear equations corresponding to physically relevant  systems such as the motion of a particle in a rotating parabola, the vibration of a tapered beam, autonomous conservative oscillator  etc. for particular choices of the parameters of the restoring force. 
Aboodh transform based homotopy perturbation method  is applied to find an approximate analytical solution to this equation giving rise to both the displacement and  frequency of the oscillation for  free vibration of strongly nonlinear oscillators as mentioned above. 
It was observed that the solution converges very fast, even first order correction is sufficient for getting results with high accuracy. 
This method  not only gives very accurate numerical values of displacement and frequency but also gives an idea about the contributions from different harmonics to it. 
It is to conclude that the solution for the generalized equation enables us to study various nonlinear physically relevant systems  easily in the same footing. The merit of ATHPM is its simplicity and ability to give the solutions to the nonlinear systems with high accuracy. This study also reveals that the ATHPM  gives  better accuracy in calculating oscillation-variables in comparison to those obtained from FAF-EBM for the systems considered.

%

\begin{thebibliography}{}
%
\bibitem{bonham1966use}
R A Bonham, and L S Su, J. Chem. Phys. {\bf 45(8)}, 2827 (1966).

\bibitem{bender1968analytic}
Carl M Bender, Tai Tsun Wu, Phys. Rev. Lett. {\bf 21(6)}, 406 (1968)

\bibitem{nayfehperturbation}
A Nayfeh, {\it Perturbation methods} ( A Willey - Interscience Publication, New York, 1973)

\bibitem{nayfehnonlinear}
A H Nayfeh and D Mook, {\it Nonlinear oscillations} (John Willey and Sons, New York 1979)

\bibitem{bogoli͡ubov1961asymptotic}
N Bogoli͡ubov, {\it Asymptotic methods in the theory of non-linear oscillations} (10, CRC Press, 1961)

\bibitem{agrwal1985weighted}
V Agrwal and H Denman, J. Sound Vib. {\bf 99(4)}, 463 (1985).

\bibitem{chen1991perturbation}
S Chen, Y Cheung and S Lau, Int. J. Nonlin. Mech. {\bf 26(1)}, 125 (1991).

\bibitem{cheung1991modified}
Y Cheung, S Chen and S Lau, Int. J. Nonlin. Mech. {\bf 26(3-4)}, 367 (1991).

\bibitem{adomian1988review}
G Adomian, J. Math. Anal. Appl. { \bf 135(2)}, 501 (1988).

\bibitem{liu1997new}
G Liu,  Conference of 7th modern mathematics and mechanics, Shanghai, 47 (1997).

\bibitem{liao1992proposed}
S J Liao, {\it The proposed homotopy analysis technique for the solution of nonlinear problems}, Ph. D. Thesis, Shanghai Jiao Tong University Shanghai (1992).

\bibitem{liu2013symbolic}
Y P Liu, S J Liao and Z B Li, J. Symb. Comput. {\bf 55}, 72 (2013).

\bibitem{he2002preliminary}
J -H He,  Mech. Res. Commun. {\bf 29(2-3)}, 107 (2002).

\bibitem{ganji2009periodic}
S Ganji, D D Ganji, Z Ganji, and S Karimpour,  Acta Appl. Math. {\bf 106}, 79 (2009).

\bibitem{he2003determination}
J -H He, Determination of limit cycles for strongly nonlinear oscillators, Phys. Rev. Lett. {\bf 90(17)}, 174301 (2003)

\bibitem{mehdipour2010application}
I Mehdipour, D D Ganji, M Mozaffari,  Curr. Appl. Phys. {\bf 10}, 104 (2010).

\bibitem{he2008improved}
J -H He, Int. J. Nonlin. Mech. Num. Simul. {\bf 9}, 211 (2008).


\bibitem{langari2011}
J Langari and M Akbarzade, Adv. Stud. Theor. Phys. {\bf 5(5-8)}, 343 (2011)

\bibitem{elnaggar2012applications}
A El-Naggar and G Ismail, Appl. Math. Sci. {\bf  6(42)}, 2071 (2012)

\bibitem{nofal2013analytical}
T A Nofal, G M Ismail, A A M Mady, S Abdel-Khalek, J. Electromagn. Anal. Appl.  {\bf 5(10)}, 388 (2013)

\bibitem{he1999homotopy}
J -H He, Comput. Methods Appl. Mech. Eng. {\bf 178(3-4)}, 257 (1999)

\bibitem{he2000coupling}
Ji-Huan He, Int. J. Nonlin. Mech. {\bf 35(1)}, 37 (2000)
\bibitem{yildirim2009homotopy}
A Y{\i}ld{\i}r{\i}m, J. Math. Phys. {\bf 50(2)}, 023510 (2009)

\bibitem{biazar2015}
Z Ayati and J Biazar, J. Egyptian Math. Soc. {\bf 23(2)}, 424 (2015)

\bibitem{biazar2011new}
J Biazar and M Eslami, Comput. Math. Appl. {\bf  62(1)}, 225 (2011)  

\bibitem{bera2012homotopy}
P Bera and T Sil, Appl. Math. Comput. {\bf 219(6)}, 3272 (2012)

\bibitem{aboodh2013new}
K S Aboodh, Global J. Pure Appl. Math. {\bf 9(1)}, 35 (2013)

\bibitem{ghorbani2009beyond}
A Ghorbani, Chaos, Solitons Fractals, {\bf 39(3)}, 1486 (2009)

\bibitem{he2008comment}
Ji-Huan He, Eur. J. Phys. {\bf 29(4)}, L19 (2008)

\bibitem{davodi2009application}
A G Davodi, D D Ganji, R Azami, H Babazadeh, Mod. Phys. Lett. B {\bf  23(28)}, 3427 (2009)

\bibitem{zhang2009periodic}
H -L Zhang, Comput. Math. Appl. {\bf 58(11-12)}, 2480 (2009)

\bibitem{marinca2006application}
V Marinca, Arch. Mech. {\bf 58(3)}, 241 (2006)

\bibitem{ganji2012energy}
D D Ganji, M Azimi, M Mostofi, Indian. J. Pure  Appl. Phys. {\bf 50}, 670 (2012)

\bibitem{billinton1985}
D. P. Billington, {\it The tower and the bridge: the new art of structural engineering} (Princeton University Press, 1985)

\bibitem{akbarzade2012dynamic}
M Akbarzade, Y Khan, Math. Comput. Model. {\bf 55(3-4)}, 480 (2012)

\bibitem{hoseini2009large}
S H Hoseini, T Pirbodaghi, M T Ahmadian, G H Farrahi,  Mech. Res. Commun. {\bf 36(8)}, 892 (2009)

\bibitem{gorman1975free}
Daniel J. Gorman, {\it Free vibration analysis of beams and shafts} (John Wiley \& Sons 1975)

\bibitem{hamdan1997large}
M Nv Hamdan, N H Shabaneh, J. Sound. Vib. {\bf 199(5)}, 711 (1997)

\bibitem{chen2009application}
S -S Chen and others, Nonlinear Anal. Real world Appl. {\bf 10(2)}, 881 (2009)

\bibitem{mathews1974unique}
P M Mathews, M Lakshmanan, Q. Appl. Math. {\bf 32(2)}, 215 (1974)

%
%
\end{thebibliography}
%

\end{document}